\documentclass[sn-basic]{sn-jnl}% Basic Springer Nature Reference Style/Chemistry Reference Style
\usepackage{graphicx}%
\usepackage{multirow}%
\usepackage[T1]{fontenc}
\usepackage{amsmath,amssymb,amsfonts}%
\usepackage{amsthm}%
\usepackage{mathrsfs}%
\usepackage{subcaption}
\usepackage{tikz}
\usepackage{makecell}
\usetikzlibrary{fit,positioning,calc}
\usepackage[title]{appendix}%
\usepackage{xcolor}%
\usepackage{textcomp}%
\usepackage{manyfoot}%
\usepackage{booktabs}%
\usepackage{lmodern}
\usepackage{hyperref}
\usepackage{algorithm}%
\usepackage{algorithmicx}%
\usepackage{algpseudocode}%
\usepackage{listings}%
\begin{document}

\title[Modal Clustering for Categorical Data]{Modal Clustering for Categorical Data}


\author*[1]{\fnm{Noemi} \sur{Corsini}}\email{noemi.corsini@phd.unipd.it}

\author[1]{\fnm{Giovanna} \sur{Menardi}}\email{giovanna.menardi@unipd.it}
%ci metteremo d'accordo su una ripartizione dei contributi che userai quando farai domanda a qualche concorso, comunque più a tuo favore

\affil*[1]{\orgdiv{Department of Statistical Sciences}, \orgname{University of Padova}, \orgaddress{\street{Via C. Battisti 241}, \city{Padova}, \postcode{35121}, \country{Italy}}}


\abstract{Despite the inherent lack of a ground truth in clustering, a broad consensus is overall acknowledged in defining the concept of cluster in the continuous setting. Conversely, this remains controversial in the presence of categorical data. We propose a novel notion of cluster based on the dual concepts of high frequency and variable association. 
We show how the concept of high frequency aligns with the cluster notion provided by modal clustering in the continuous setting, which allows us to borrow and adapt existing operational tools to develop a novel procedure. The method is illustrated on some real data and tested via
simulations.}




\maketitle

\section{Introduction}

%L'idea è partire dal dato categoriale e proporre un metodo di clustering per quello, basato su un concetto di cluster ben definito, poi toh guarda caso il concetto di cluster che sviluppiamo assomiglia a quello modal, e quindi operativamente sfrutteremo gli strumenti modal --> quindi partirei con un ampliamento dell'intro cladag (ampliamento $\neq $ aggiungo solo chiacchiere di contorno per renderlo più lungo). 
%Occhio a non partire dalla rava e la fava per introdurre il clustering: in un paper scientifico chiunque sa cos'è il clustering
%In sostanza: mix tra intro cladag e 1.3.1 tesi + summary esaltare il fatto che il metodo determina automaticamente il numero di gruppi
The central role of clustering in statistics has never been questioned over the years, thanks to its many relevant applications across various fields. Consider, for example, market analysis with customer segmentation, the classification of genes in biology, or recommendation systems in information technology, among many others. Beyond its wide applicability, a further reason for the proliferation of a voluminous amount of literature on this topic can be traced back to the inherent unsupervised nature of the clustering problem. The lack of a unique and unambiguous, formally sound ground truth to aim at has favoured the spreading of numerous declinations of clustering concepts and associated methods to identify them.

Nevertheless, when numerical data are at hand, a general agreement is met across alternative notions of cluster, which collectively fall under the heading of groups of similar subjects. Even when more sophisticated density-based cluster formulations are considered, indeed, the underlying notion of cluster implies the observations to be somewhat close to each other. 
Conversely, this is heavily hindered with categorical data. Here, in principle, a natural clustering gathers subjects within the observed cross-categories of the variables. However, such description turns out to lack parsimony when either the number of variables and/or the number of categories grows. On the other hand, the lack of a total order among categories makes somewhat controversial even the notion of distance, and introduces a degree of arbitrariness in aggregating cross-categories to form clusters \citep*[see, e.g.,][]{boriah2008similarity, van2024general}. Indeed, the notion of cluster itself is usually left unspecified in the inherent literature, and implicitly follows from the choice of the applied clustering method, rather than driving the development itself of methods pursuing a well-defined population goal.   
%In fact, even within the setting of categorical data, we may formalise some concept of cluster building on some natural intuition. 

In this paper, we attempt the ambitious aim of proposing a novel notion of cluster within the setting of categorical data, along with an associated operational procedure to identify groups. The proposed notion of cluster is based on the dual concepts of high frequency and association between variables and notably complies itself with some intuition of groups formed by categorical data. We show how the concept of high frequency aligns with the cluster definition provided by modal clustering \citep[see, e.g., ][]{menardi2016review}, which allows us to borrow and adapt existing operational tools to develop a novel procedure to identify groups. The proposed procedure notably 
%The proposed methodology not only aligns with this intuitive notion of clustering but also offers a distinct advantage: it 
does not require the number of clusters to be specified in advance, a limitation present in most methods available in the literature. 

The rest of this paper is organized as follows: Section \ref{background} provides an overview of the
building blocks needed for the subsequent developments, by reviewing the literature on clustering categorical data, and outlining the key notions on modal clustering for continuous data. Section \ref{MCM} defines the novel concept of clusters for categorical data and presents the proposed method, which is afterwards discussed in Section \ref{modal_discussion}.
Section \ref{numerical_studies} presents a
simulation study and some real data applications. Finally, Section \ref{conclusion} provides concluding remarks.

\section{Background}
\label{background}
\subsection{Related works}
\label{related_works}

%The ill-posedness of the clustering problem, and the subsequent proliferation of a plethora of heterogeneous methods to address it, hold particularly true in the categorical data setting, where the lack of a total order among categories results in a certain degree of arbitrariness even in the concept of distance itself.
Due to the reasons outlined above, overviewing a taxonomy of the existing methods to cluster categorical data is challenging. Nevertheless, similarly to the continuous setting, the majority of the approaches can still be broadly framed into distance- and model-based approaches.

Within the former context, relationships between observations are quantified by gauging either dissimilarities or distances. The choice of a specific metric defines how observations are considered similar, with the most common in the literature being the the simple matching distance, the Hamming distance, and the Jaccard distance. 
One of the most commonly used methods in this framework is the Partitioning Around Medoids (PAM) method developed by \citet{kaufman2009finding}.
PAM seeks to identify $K$ representative observations, known as medoids, to which observations are assigned, aiming to minimize the sum of dissimilarities or distances. See \citet{vrezankova2009cluster} for a comprehensive review of metrics for categorical data. 

As to the model-based approach, Latent Class Analysis (LCA) is perhaps the most widely studied and used method for clustering categorical data.
Introduced by \citet{vermunt2002latent}, it models data as arising from a mixture of multinomial distributions, with each cluster corresponding to a single mixture component, called latent class, defined by a probability distribution. The contingency table can be modelled as a finite mixture of unobserved tables generated under the assumption of conditional independence with respect to the latent variable. The variables are therefore considered independent conditionally on the clustering structure.
Recently, an interesting model-based clustering method for categorical data, called \texttt{clustMD}, has been introduced by \citet{mcparland2016model}. Designed for mixed-type data, this method can handle both continuous and categorical data separately. It assumes that the observed variables are manifestations of an underlying continuous latent vector, distributed according to a mixture of Gaussian distributions, which produces the observed mixed-type data. By specifying a particular structure for the covariance matrix, various clustering models can be delineated. For further details, refer to \citet{mcparland2016model}.

Within the modal setting, a few methods have been proposed, and most still rely on some concept of distance. One of the most well-known methods in this context is the $K$-modes algorithm by \citet{huang1998extensions}, an extension of the renowned $K$-means algorithm, that relies on the concept of mode.
Other noteworthy approaches for clustering categorical data in the non-parametric framework include the work by \citet{chen2016clustering} and the method proposed by \citet*{beck2021new} which emulates, somehow, the idea of gradient-ascent methods for continuous data, based on the evaluation of the observations neighborhood. 
The only attempt to closely rely on a sort of density measure for categorical data was made by \citet{giordan2011clustering}, which in fact handle ordinal data only.

\subsection{Modal clustering of continuous data}

The rationale underlying the modal approach to clustering lies in formalizing the intuitive concept of clusters as dense sets of observations. This is achieved within a rigorous framework that connects the notion of groups to specific features of the probability density function assumed to generate the data. In essence, there is a one-to-one correspondence between clusters and modal regions of the density, where the modes represent the archetypes of the clusters. An illustrative analogy comes from terrain analysis: if the density is figured as a mountainous landscape, with the modes as peaks, a modal cluster is the region that would be flooded by water flowing from a fountain placed at a peak. \citet{chacon2015population} formalized this idea by means of Morse Theory \citep[see][for an introduction]{matsumoto2002introduction}, a branch of differential topology describing the large scale structure of an object via the analysis of the critical points of a function.

\begin{figure}[t]
	\centering
	\begin{minipage}{.33\textwidth}
	    \centering
		%\textsf{(a) }\par\medskip
		\includegraphics[width=\linewidth, height = 5cm]{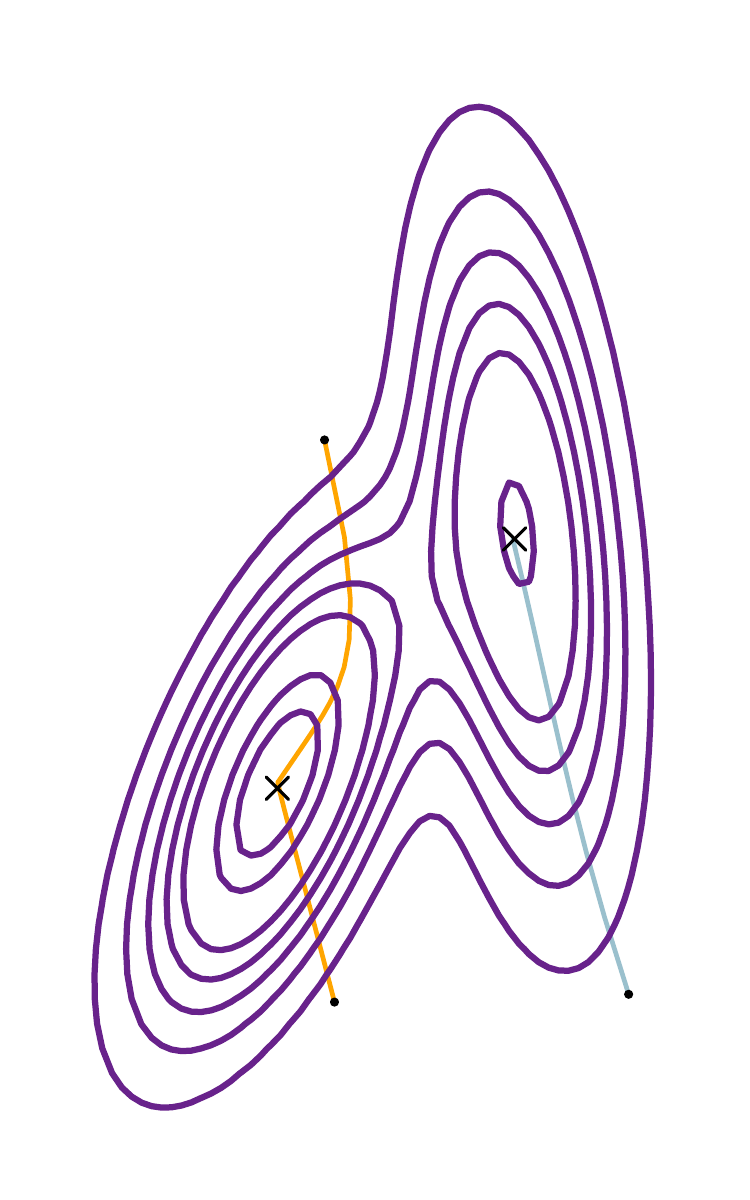}
		%\captionof{figure}{Cartography of Switzerland}
	\end{minipage}\hfill
	\begin{minipage}{.33\textwidth}
		\centering
		%\textsf{(b) }\par\medskip
		\includegraphics[width=\linewidth, height = 5cm]{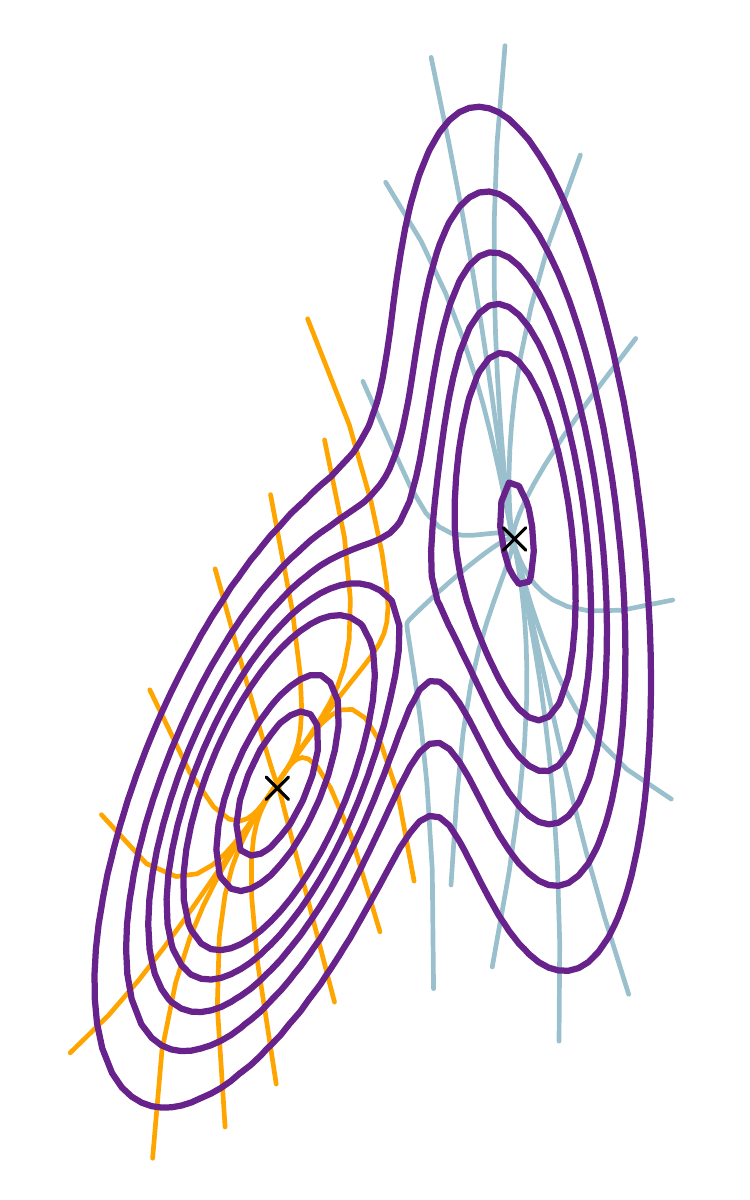}
		%\captionof{figure}{Mesh}
	\end{minipage}\hfill 
	\begin{minipage}{.33\textwidth}
		\centering
		%\textsf{(b) }\par\medskip
		\includegraphics[width=\linewidth, height = 5cm]{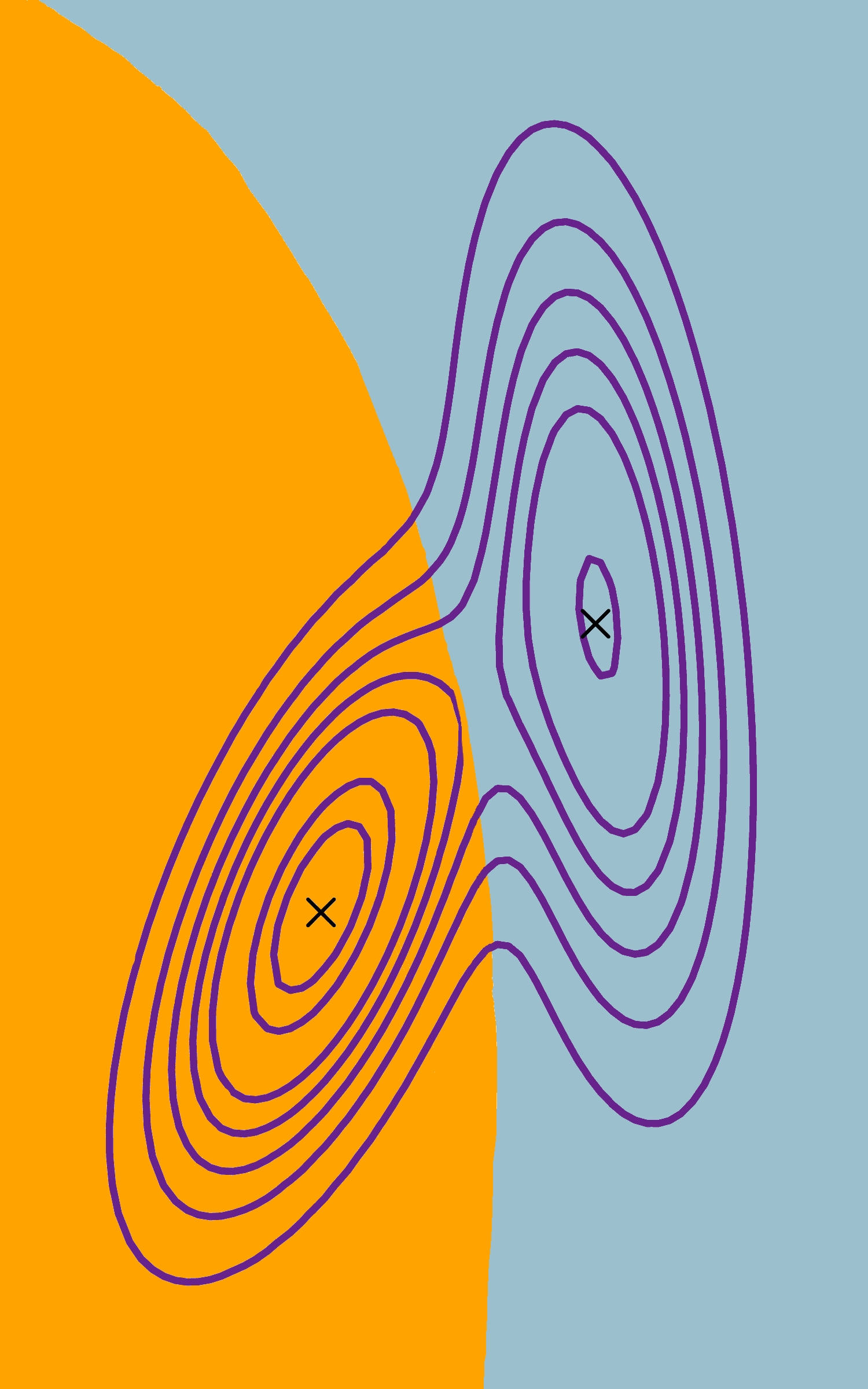}
 %           \includegraphics[width=\linewidth, height = 5cm]{images/ex2ms.jpeg}
		%\captionof{figure}{Mesh}
	\end{minipage}
	\caption{An example of bimodal $f$ with the integral curves of the negative gradient starting from some random points (left and middle panel) and the partition induced by the destination of the integral curves in basins of attraction of the modes of $f$ (right panel).}
	\label{fig:idealmodalpop}
\end{figure}

Formally, let $X$ be a random vector, with probability density function $f\colon\mathbb{R}^d \rightarrow \mathbb{R}$. Assume  $f$ to be a Morse function, i.e. a (smooth) function with no degenerate critical points, and denote by $\theta_1,\dots, \theta_M$ the modes of $f$ (i.e. its local maxima). For a given $x \in \mathbb{R}^d$, define the \emph{integral curve} of the gradient $\nabla f,$ as the uphill path $\nu_x : \mathbb{R} \rightarrow \mathbb{R}^d$, such that
%\begin{eqnarray*}
$\nu'_x(t) = \nabla f(\nu_x(t)),$ with $\nu_x(0)=x.$
%\end{eqnarray*}
The \emph{domain} of attraction of a critical point $x_0$ is defined as the set of points whose integral curve eventually converges to $x_0$ (as $t\rightarrow \infty$)
$$ \mathcal{D}(x_0) = \{ x \in \mathbb{R}^d : \lim_{t\rightarrow  \infty} \nu_x(t)=x_0  \}.$$
With these notions at hand, the modal population clustering $\{ \mathcal{D}(\theta_1),\dots,\mathcal{D}(\theta_M) \}$ associated with $f$ is described by the set of the domains of attraction of the modes of $f,$ which yields a partition of the whole space.
%By borrowing concepts from terrain analysis, the underlying intuition is that, if $f$ is figured as a mountainous landscape where the modes are the peaks, a modal cluster is the region that would be flooded by a fountain emanating from a peak. %When $d=1$, clusters are then unequivocally defined by the locations of the minima points of $f$, which represent the cluster boundaries. 
An illustration is provided in Figure \ref{fig:idealmodalpop}. 

Since the true density $f$ is unknown, a nonparametric estimator $\hat f$ is typically used to ensure flexibility in identifying clusters of arbitrary shapes. A common choice is the kernel density estimator \citep[see][for a detailed review]{chacon2018multivariate}. %, defined as: \begin{eqnarray}\label{eq} \hat f(x) = \frac{1}{n} \sum_{i=1}^{n} K_H(x - x_i), \end{eqnarray} where $H$ is a bandwidth matrix and $K(\cdot)$ is a smooth, symmetric kernel function.
Once the density function has been estimated, to identify operationally the modal regions of $\hat f,$ methods performing modal clustering mostly follow two main strands. 

A first strand takes its steps from the mean-shift algorithm \citep{fukunaga1975estimation}, essentially a variant of the gradient ascent algorithm which discretizes the concept of integral curves converging to the modes by iteratively shifting points toward the local maxima of $\hat f$ via the steepest ascent path. Starting from an initial point $x^{(0)}$, a sequence of steps is built according to the following update mechanism: 
\begin{equation*}\label{eq}
x^{(r)} = x^{(r-1)} + A \frac{\widehat{\nabla f}(x^{(r-1)})}{\hat f(x^{(r-1)})} 
\end{equation*} 
where $A$ is set to ensure convergence. This process continues until convergence to a mode.  
The final partition is obtained by grouping all points that ascend to the same mode.

Alternatively, modal regions are associated to disconnected density level sets of the
sample space, without attempting explicitly the task of mode detection. When there is no clustering structure, $f$ is unimodal, and any section of
$f$, at a given level $\lambda$, singles out a connected (upper) level set: $L(\lambda) = \{x \in \mathbb{R}^d : f(x) \geq \lambda\}$. Conversely, when $f$ is multimodal, $L(\lambda)$ may be either connected or disconnected,
depending on $\lambda$. In the latter case, it is formed by a number of connected components,
each of them associated with a region of the sample space including at least one mode
of $f$. Since a single section of $f$ could not reveal all the modes,  $\lambda$ is moved along its
feasible range, giving rise to a hierarchical structure, known as the cluster tree, which
provides the number of connected components of $L(\cdot)$ for each $\lambda$. Each leaf of the tree describes
a cluster core, defined as the largest connected component of the density level set which
includes one mode. Operationally, $f$ is replaced by $\hat f$ and the connectedness of the density level sets is evaluated by means of graph theory.

\section{Modal clustering of categorical data}
\label{MCM}

\subsection{A novel notion of cluster for categorical data}
\label{novelnotion}

\begin{figure}[b]
	\centering
	\includegraphics[height=5cm]{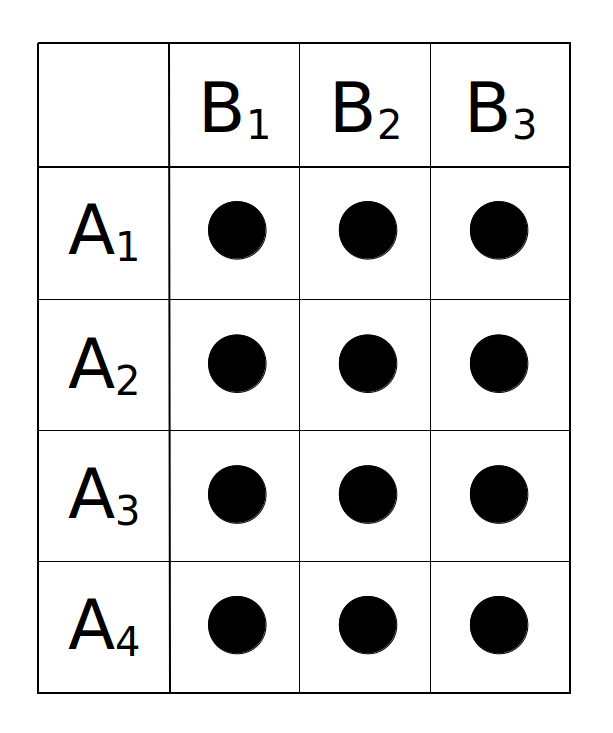}	
	\includegraphics[height=5cm]{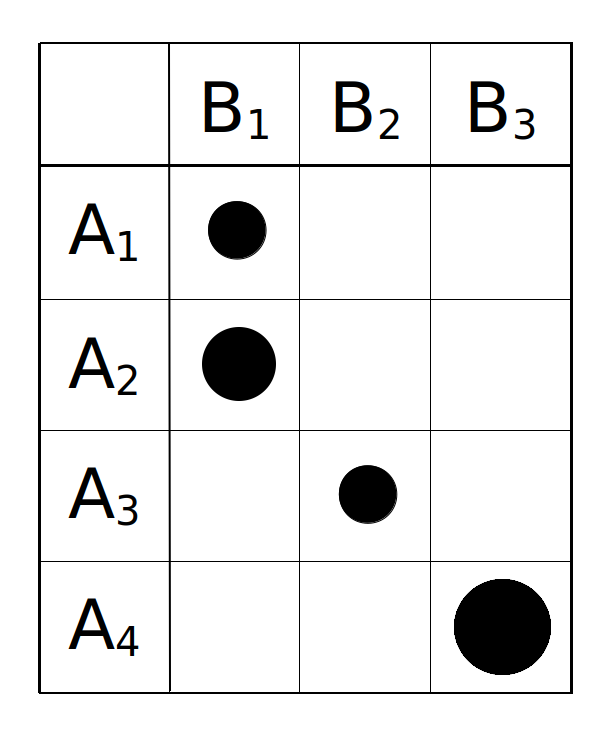}
	\caption{\label{fig:toy} Graphical representation of two contingency tables, with two different frequency patterns, where the diameter of each circle is set as proportional to the frequency of the cell. The right example conveys the idea of clustered subjects, while this does not occur in the left example.}
\end{figure}

To gain a first insight into the concept of a cluster we will refer to, consider the two toy examples in Figure \ref{fig:toy}, which describe two alternative frequency patterns of a number of subjects observed with respect to two variables. According to our perception, and even in the lack of a cluster definition, a configuration without clusters (or formed by 12 clusters, as the number of
cross-categories) appears in the left panel, where variables are independent; conversely, in the right panel, characterized by a strong
association between the variables, the frequency pattern suggests the aggregation of the subjects into two or three
clusters.

To some extent, similarly to the modal formulation of the cluster concept, the most populated cells act as anchors
to attract observations and form clusters. However, with respect to the continuous setting, where the notion of mode just refers to a locally high-density point, in the categorical setting the same notion should be intended as an excess probability mass compared to the situation of mutual independence between variables.
This intuition leads us to build a novel notion of cluster hinging on the twofold concept of high frequency and association between 
variables.  More formally, given two variables $A$, $B$, with levels $\{A_1, \ldots, A_M\},$ and $\{B_1, \ldots, B_L\}$ respectively, the modes will be identified among the cross-categories $(A_m, B_l)$ for which the ratio
\begin{equation}
    \label{eq:assoc}
\frac{P(A = A_m, B = B_l)}{P(A = A_m)P(B= B_l)}
\end{equation}
is high. On the other hand, the concept of domain of attraction which, in the continuous setting, allows for partitioning the observations depending on the pertaining mode, requires the definition of some aggregation criterion. Here again, evaluating the association between the observed variables comes to the aid, as different cross-categories will make sense to belong to the same cluster when they hold similar information. Hence, the cross-categories $(A_m, B_l)$ and $(A_{m'}, B_l)$ shall be considered as highly connected not only because they share the same level for variable $B$ but also when, given that $B=B_l$, both $A_m$ and $A_{m'}$ become more likely, that is, when 
\begin{equation*}%\label{eq:pcond_ratio}
\frac{P(A_m|B_l)}{P(A_m)} \quad \mbox{ and } \quad \frac{P(A_{m'}|B_l)}{P(A_{m'})}
\end{equation*}
are high, which occurs when the respective \eqref{eq:assoc} are also jointly high.

\subsection{The proposed method}
\label{sec:novelmethod}

%Given the conceptual analogy between the notion of modal clusters in the continuous setting and the notion of clusters for categorical data outlined above, we shall also borrow from the continuous setting some operational tools  to define a procedure for identifying clusters in categorical data.

In the continuous setting, modal regions are operationally identified either as the set of points whose direction of the steepest gradient ascent path converges to the same mode, or as connected sets with density above a threshold.
In the categorical setting, defining both a density or its gradient is precluded and, at the same time, there is no obvious method to define the connectedness of a region. Nevertheless, we shall define a procedure that jointly extends both these ideas, if not formally, at least conceptually, based on the principles outlined above and building on some notions from graph theory.    

Consider a sample of $N$ subjects, observed, for the sake of simplicity, with respect to two categorical variables $A \in \{A_1, \ldots, A_M\}$, and $B \in \{B_1, \ldots, B_L\}$. The generalization for a larger number of variables will be discussed later. It is convenient to move from the sample space of the observed pairs to the contingency table providing for each cross category $(A_m, B_l)$ its joint frequency $N_{ml}$, $m=1, \ldots, M,$ $l=1, \ldots, L$.

%$V_i = (A_m, B_l), i= 1, \ldots, M\times L$
%With the aim of defining a procedure that jointly extends both the level set and the mean-shift techniques to categorical data, we build a directed weighted graph with the vertices being the cross-categories of the contingency table.
%To achieve this objective, it is essential to define a notion of density, from which a notion of mode can be derived, a notion of connectedness within the graph, and a path addressing the mode.

Following the intuitions outlined in Section \ref{novelnotion}, we can define a measure of density based on the node’s attractiveness, which accounts for high frequency and association between variables. To this aim, we may in principle stem from any measure of association between categorical variables. A natural choice, for example, would select the individual contributions to the chi-square measure, or to its normalised version Cramer’s V \citep[see, e.g.,][]{agresti2012categorical}. In fact, we build on the \emph{mutual information} \citep[MI,][]{shannon2001mathematical} defined as the Kullback-Leibler divergence between the joint distribution of $(A, B)$ and the expected distribution under the assumption of independence: 
%Let $(A, B)$ be a pair of categorical variables with $m = 1, \dots, M$ and $l = 1, \dots, L$ possible levels, respectively. Then, it is possible to express their mutual information $MI(A, B)$ as
\begin{equation*}
	\begin{aligned}
	MI(A, B) &= H(A) + H(B) - H(A,B) =\\
	&=  \sum_m \sum_l P(A= A_m, B = B_l) \log \left[\frac{P(A = A_m, B = B_l)}{P(A = A_m) P(B = B_l)}\right]
	\end{aligned} 
\end{equation*}
where $H(A) = - \sum_{m=1}^M P(A=A_m)\log P(A=A_m)$ is the entropy of $A$. This measure quantifies the \textit{amount of information} on one variable held by the other one. Its values range from $0$, occurring when the variables are independent, to the entropy of $A$, occurring when $B$ is a deterministic function of $A$, or viceversa. 
It is worth noting that, in case of independence, this value is equal to $0$ even if the joint probability $P(A= A_m, B = B_l)$ is large. Higher MI indicates greater clusterability of the data. 

As a measure of density of each cross-category, we then consider its individual contribution to the mutual information, namely
\begin{equation}
\label{eq:mutual_info}
	\delta\left(A_m, B_l\right) \propto P\left(A = A_m, B = B_l\right) \log \left[\frac{P\left(A = A_m, B = B_l\right)}{P\left(A = A_m\right) P\left(B = B_l\right)}\right]
\end{equation}
afterwards normalised by min-max normalisation.
Note that Equation \eqref{eq:mutual_info} satisfies both the requirements of depending on the probability of occurrence and on the amount of information that overlaps between variables. 
%The importance of each node is assessed not only by considering its observed frequency but also by quantifying the extent to which the observed frequency of a cross-category exceeds the frequency expected under the assumption of independence of the variables, i.e., the frequency that we would expect a priori in a random scenario. 
By normalising it, its values range between $0$ and $1$, with $0$ indicating that the cross-category provides no additional information beyond that discernible when the variables are independent. Conversely, values greater than $0$ are intended as the amount of overlapping information held by $A$ and $B$ and due to $(A_m, B_l)$. Furthermore, by normalizing the individual contribution to the mutual information, we no longer need to distinguish between positively or negatively associated cross-categories. As a result, the higher the node's normalised individual contribution, the greater its attractiveness. %The concept of cluster is not uniquely based on high probability but rather it focuses on the discrepancy between the observed probability and the expected probability under the assumption of independence.
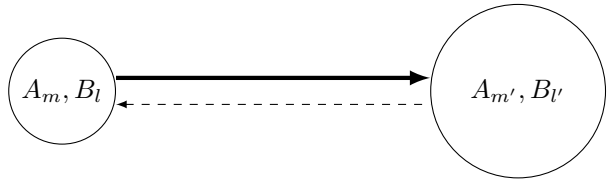
\begin{figure}[b]
\centering
    \begin{tikzpicture}
      \node[minimum width=1cm,draw,circle] (a) at (0,0) {$A_m, B_l$};
      \node[minimum width=2.3cm,draw,circle] (b) at (6,0) {$A_{m^\prime}, B_{l^\prime}$};
      \draw[>=latex,->, ultra thick] ([yshift= 5pt] a.east) -- node[above, draw=none, rectangle]{} ([yshift= 5pt] b.west);
      \draw[>=latex,<-, thin, dashed] ([yshift= -5pt] a.east) -- ([yshift= -5pt] b.west);      
\end{tikzpicture}
\caption{Edge weight and direction of a pair of nodes. The diameter of each circle is proportional to the normalised contribution to the mutual information}
\label{fig:node_direction}
\end{figure}

Once a density measure is defined, resorting to graph theory and building a suitable directed and weighted graph allows us to translate the concepts of gradient ascent, and equivalently the connectedness of a region, needed to define the domain of attraction of a mode, into the categorical setting. Specifically, each cross-category $(A_m, B_l)$ is associated with a node $V_i$, $i = 1, \dots,M\times L$, and a link between two nodes sharing one category is set with weight equal to the minimum density, in the direction of the node with maximum density (Figure \ref{fig:node_direction}). 

Formally, given $V_i = (A_m, B_l)$ and $V_{i^\prime} = (A_{m'}, B_{l'})$, the weight $\omega_{i, i^\prime} $ between the two nodes will be

\begin{equation}
\label{weights}
\omega_{i, i^\prime}  = \left\{
\begin{array}{ll}
\min(\delta(A_{m}, B_{l}), \delta(A_{m'}, B_{l'}))& \mbox{if }m=m' \mbox{ or } l=l'\\
0 & \mbox{otherwise.}
\end{array}
\right.
\end{equation}

This guarantees that the strength of the connection between $(A_m, B_l)$ and $(A_{m'}, B_{l'})$ will be high when, given that $B=B_l$, and $l={l'},$ both $A_m$ and $A_{m'}$ become more likely, and the occurrence of both the events $(A_m, B_l)$ and $(A_{m'}, B_{l'})$ is high. 

\begin{figure}[b]
	\centering
	\begin{tabular}{cc}
	\includegraphics[height=5cm]{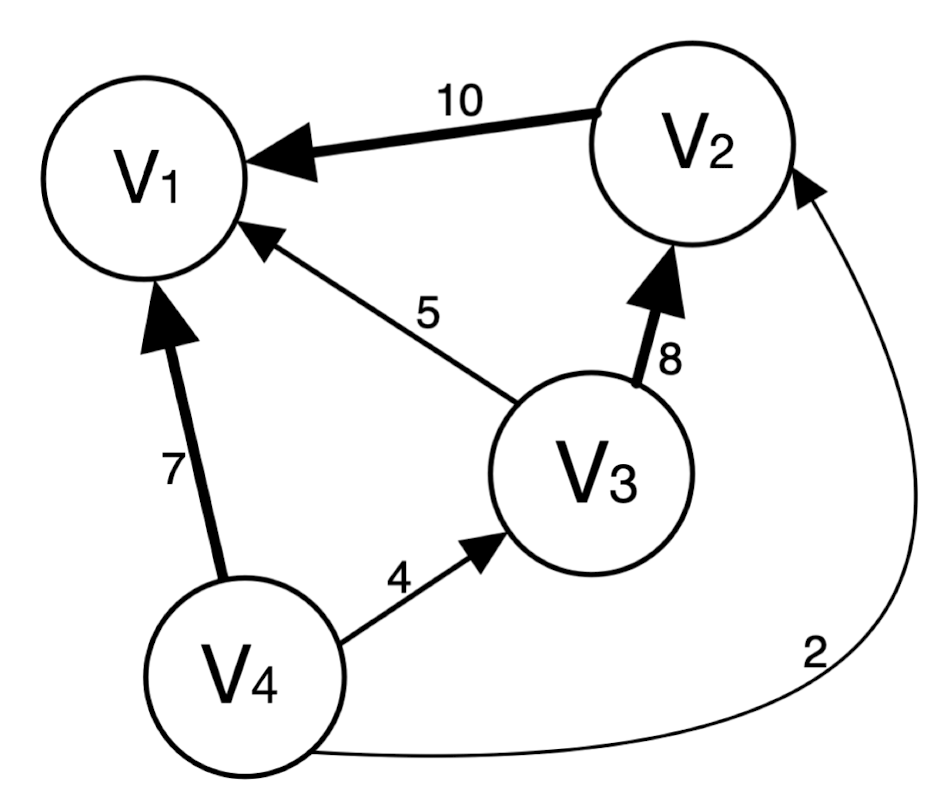} &
	\includegraphics[height=5cm]{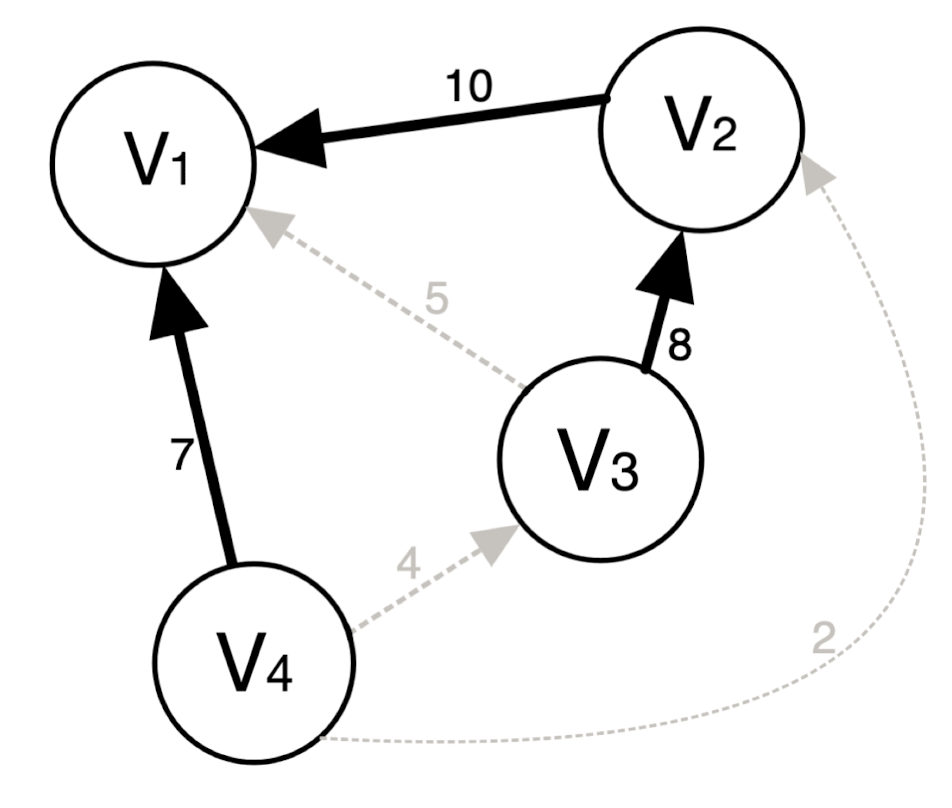}\\
	\end{tabular}
	\caption{\label{fig:mst} Maximal spanning tree (right) obtained starting from a toy graph (left). Values on each edge represent (examples of) their weight. For nodes with multiple outgoing edges, the MST algorithm ensures that only the highest-weight edge is retained.}
\end{figure}
Once the graph is built, its maximal spanning tree (MST) is derived to trim  multiple paths from the graph, ensure that each node is restricted to one outgoing edge only, and that the resulting graph has maximum weight. To this purpose, the Chu-Liu/Edmonds algorithm \citep{chu1965shortest, edmonds1967optimum}, in principle aimed to find the minimum spanning tree, can be suitably adjusted, as designed for directed graph. See Figure~\ref{fig:mst} for a simple illustration.
Given that the initial graph is not fully connected, the MST may result in multiple subgraphs. Specifically, a collection of $K$ maximal spanning trees, known as a maximal spanning forest, is obtained. Each element of this forest represents a group of cross-categories, eventually associated with a cluster. 

In addition to allow for the identification of both the composition and the number of groups, the discussed way of proceeding provides a twofold lens: on one hand, similarly to mean-shift type algorithms, within each subgraph, a path originates from every node, following the direction of steepest ascent and with the final destination being the mode (namely, the cross-category with locally maximum density, according to definition \eqref{eq:mutual_info}); on the other hand, each subgraph enables the identification of groups as connected components of elements with density higher than a given threshold, and by varying this threshold, it is possible to define a hierarchy of clusters in the form of a tree. 
%This results in a path that addresses the locally most attractive node. The idea of the steepest gradient ascent path is therefore translated into a sequence of links between nodes driving in the direction of the node with the locally highest contribution to mutual information. 
%The final result can be simultaneously visualised and interpreted from both a mean-shift and level-set perspective, proving, at least empirically, that even for categorical data, level-set and mode-shift results are equivalent, as previously established in the continuous case by \cite{arias2023moving}. From the mean-shift perspective, each cluster is seen as a sequence of links leading towards the node with the locally highest contribution to mutual information. From the level-set perspective, each cluster is a set of connected nodes with a locally highest contribution to mutual information above a certain threshold.
Algorithm \ref{alg:one} outlines the relevant steps of the introduced clustering procedure, in the following referred to as \emph{modal clustering of categorical data} (MC-C).

\begin{algorithm}[t]
\caption{\label{alg:one} MC-C: Modal Clustering for Categorical data}
\begin{algorithmic}[1]
\State \textbf{Input}: Set $\{c_1, \dots, c_N\}$ with $c = (A,B)$, $A \in \{A_1, \dots, A_M\}$, $B \in \{B_1, \dots, B_L\}$
\State \textbf{Obtain the cross frequency table}
\State \indent $N_{ml} = \#(A_m, B_l)$ with $m = 1, \dots, M$ and $l = 1, \dots, L$
\State \textbf{Build a directed weighted graph} $\mathcal{G} = \{V_i, \omega\}$ 
\State \indent Assign a cross category to each node $V_i = (A_m, B_l), i = 1, \dots, M\times L$ 
\State \indent Based on $N_{ml}$ determine the density of each node $\delta(V_i) = \delta(A_m, B_l)$ as in Equation~\ref{eq:mutual_info}
\State \indent Apply the min-max normalisation to the computed densities
\State \indent \textbf{for} each pair $(V_i$, $V_{i^\prime}$) \textbf{do}
    \State \indent \indent Compute the weight for the link $\omega_{i, i^\prime}$ as in Equation \ref{weights}
    \State \indent \indent Point the link towards the node with the maximum density
\State \indent \textbf{end for}
\State \textbf{Determine the maximal spanning forest of} $\mathcal{G}$
\State \indent Apply the Chu-Liu/Edmonds algorithm and find MST($\mathcal{G}$)
\State \textbf{Determine} $\mathbf{K}$ \textbf{connected components of MST(}$\mathcal{G}$\textbf{)}, $\mathcal{G}_1, \dots, \mathcal{G}_K$
\State \indent Assign $c = (A, B)$ to $g_k$ if $V_i = (A, B) \in \mathcal{G}_k$
\State \textbf{Output}: Cluster labels $g_1, \dots, g_N$, with $g_n \in \{\mathcal{G}_1, \dots, \mathcal{G}_K\}$
\end{algorithmic}
\end{algorithm}

\begin{figure}[tb]
    \centering
    \begin{tabular}{ccc}
    	\includegraphics[width=0.15\textwidth]{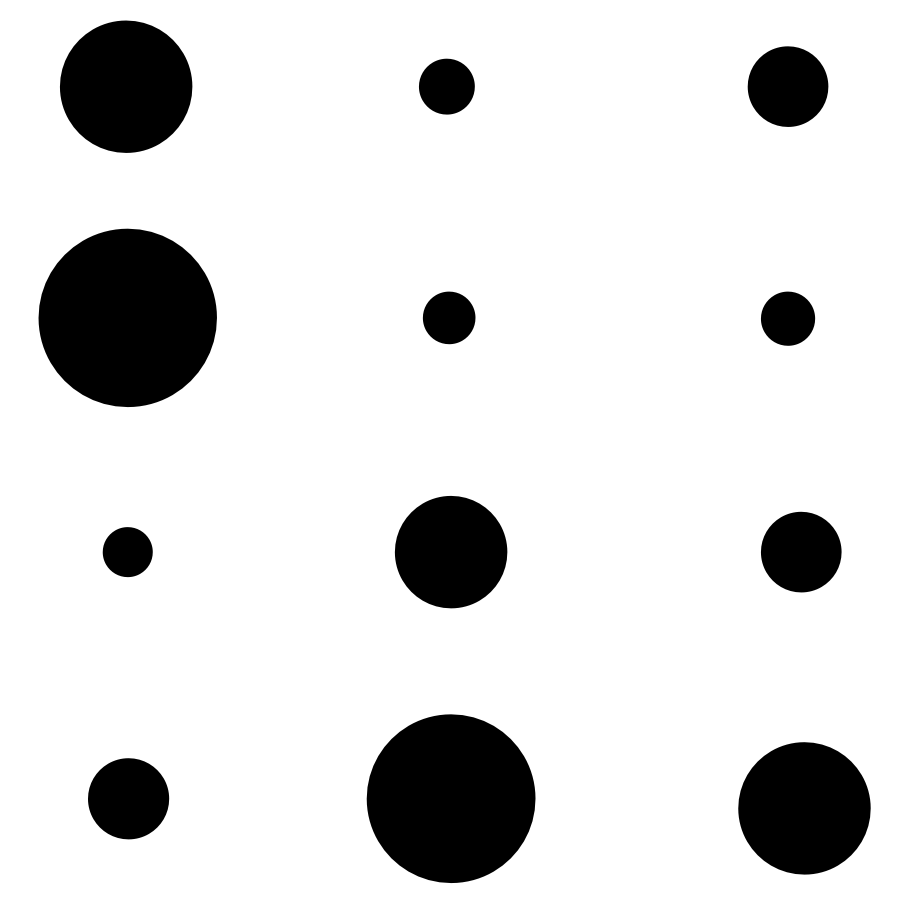} &
    	\includegraphics[width=0.2\textwidth]{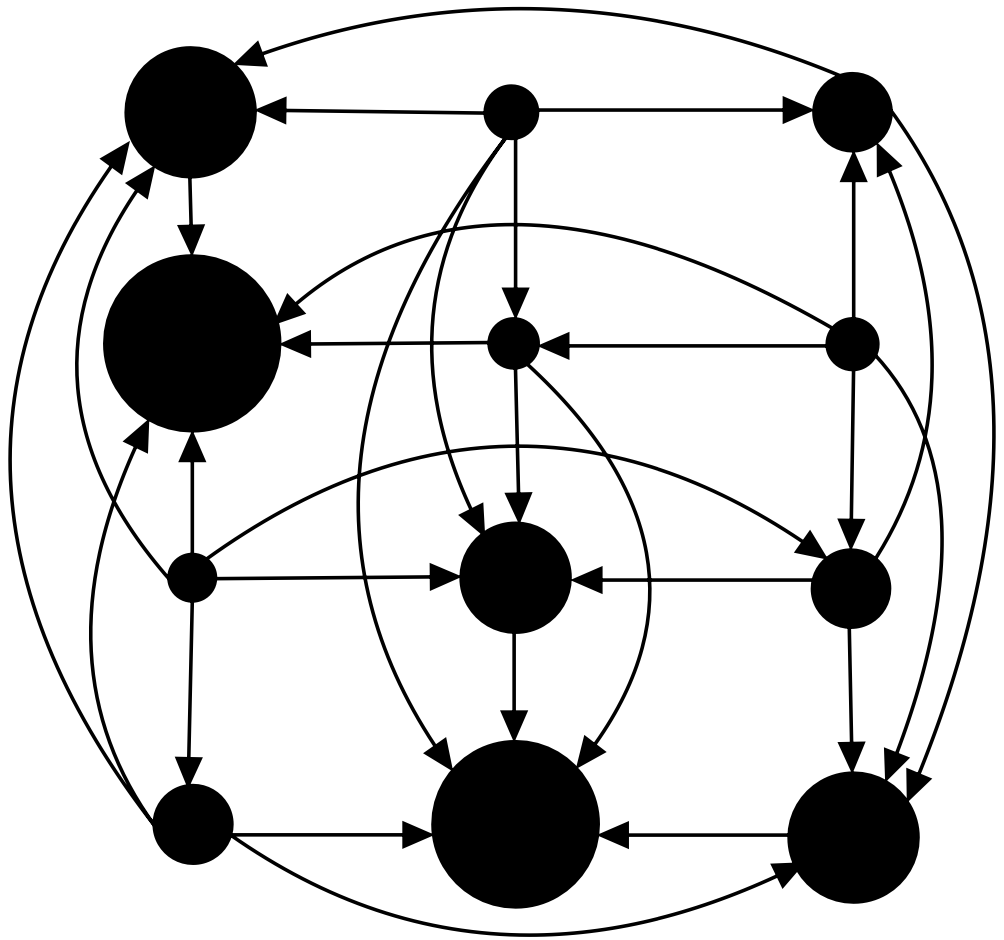} &
    	\includegraphics[width=0.2\textwidth]{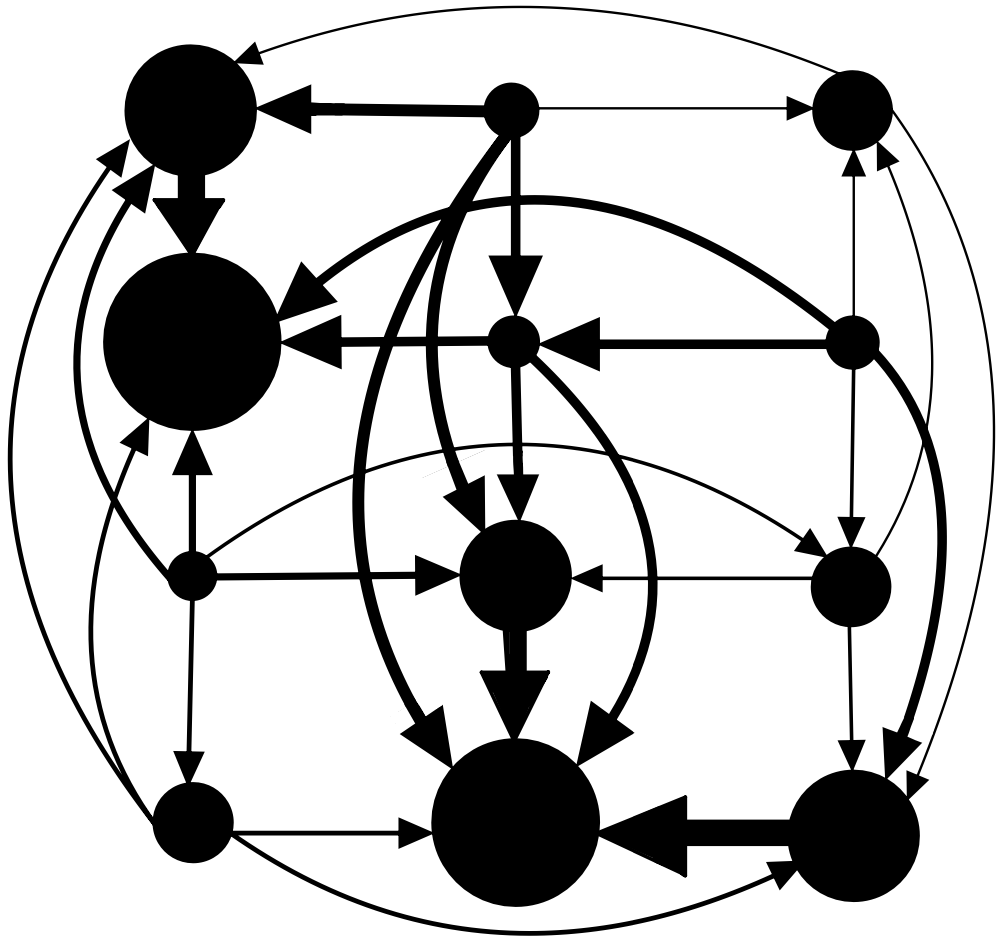}\\
    	(A) & (B) & (C) \\
    \end{tabular}
    \begin{tabular}{ccc}
    	\includegraphics[width=0.2\textwidth]{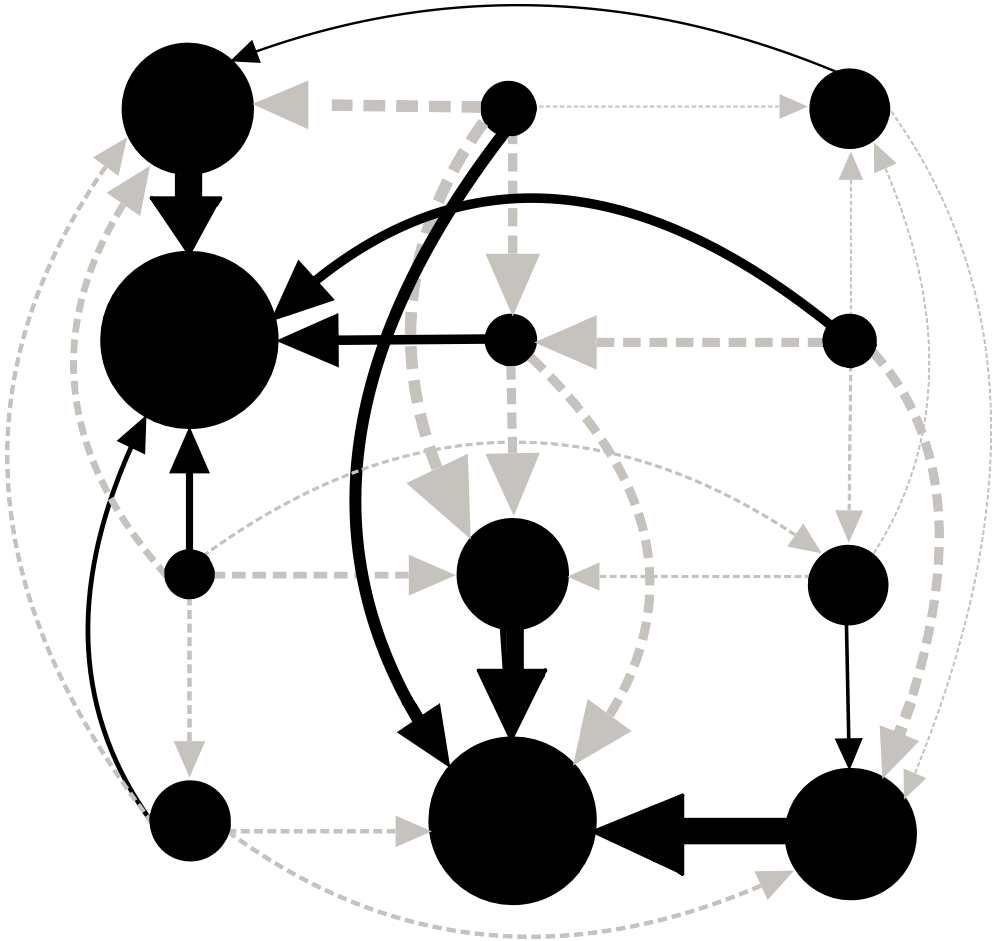} &
    	\includegraphics[width=0.2\textwidth]{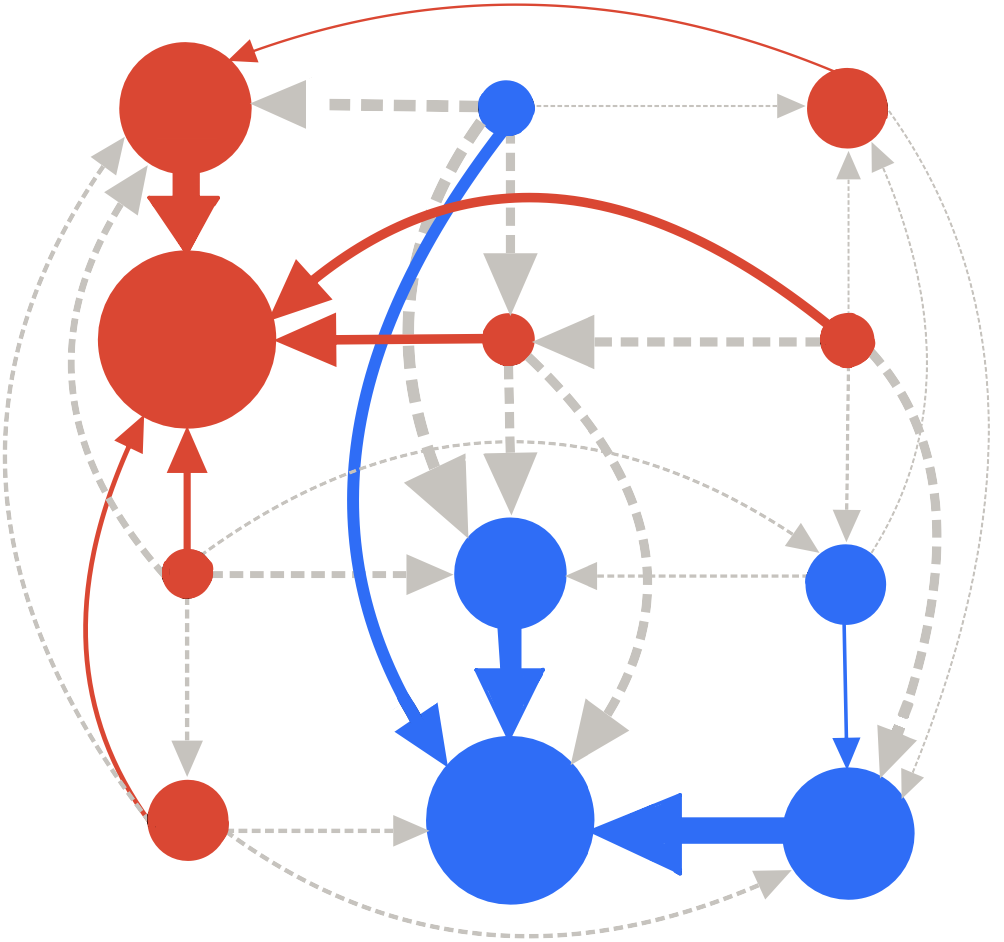} & 
    	\includegraphics[width=0.2\textwidth]{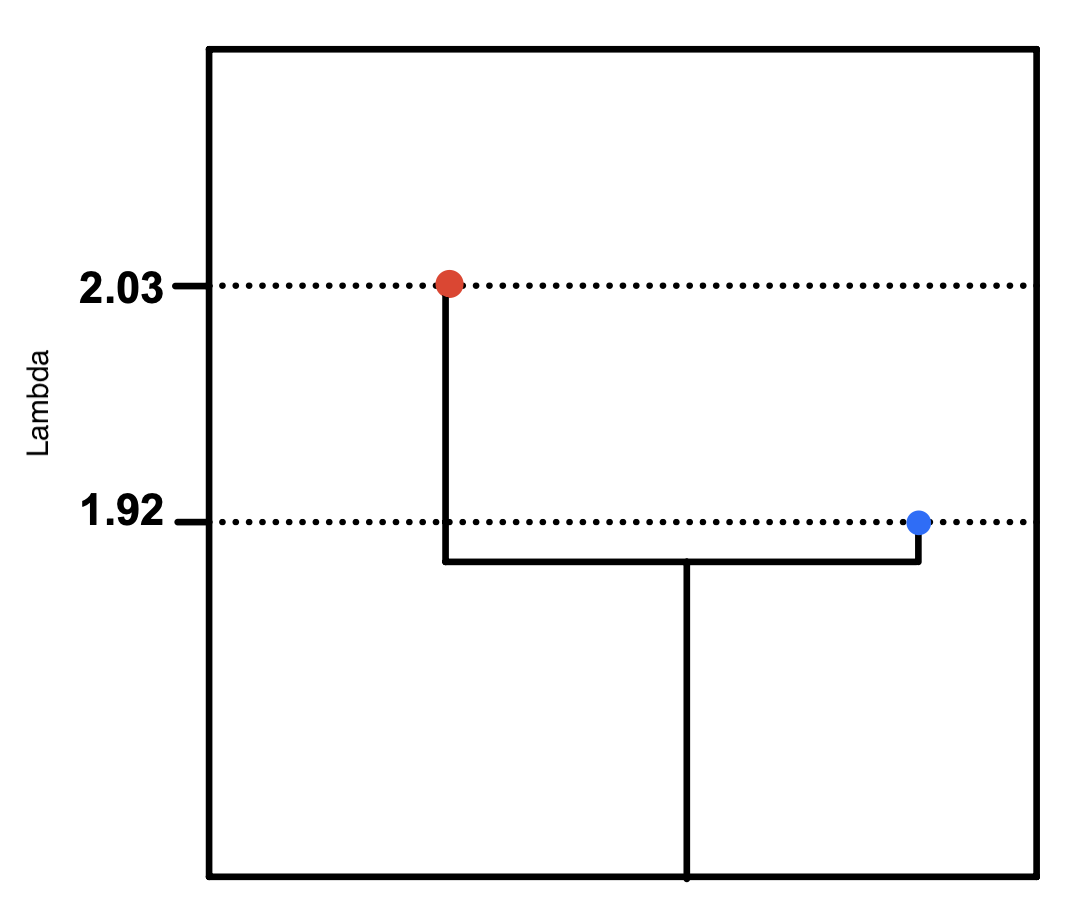}\\
    	(D) & (E) & (F)\\
    \end{tabular}
    \caption{Step-by-step procedure for implementing the proposed clustering method (A-E) and the obtained cluster tree (F). It begins with the construction of the graph (A, B, C), progresses through finding the MST (D), and concludes with the final clustering outcome (E) and the visualisation of the cluster tree (F).}
    \label{fig:methodoperationally}
\end{figure}

An illustration of these steps, referred to the Example in  Table~\ref{tab:illustrative}, is provided in Figure \ref{fig:methodoperationally}. A sample of 
 $462$ individuals are cross-classified according to the variables \textit{religion} and \textit{geographic area of origin}. By examining the observed frequencies, the two variables clearly exhibit a high dependency structure. 
 From this contingency table, the graph is build, as showed in Figure \ref{fig:methodoperationally}, by assigning each cell to a node (A), and setting a link between any two nodes that share at least one level. Links are directed towards the node with the higher normalised individual contribution to the mutual information, as indicated by the node sizes (B). The weight of each link is set to the minimum of the two aforementioned contributions (C). The maximal spanning tree of the graph is then computed (D), resulting in two disconnected subgraphs that correspond to the clusters (E). 
One group, led by the USA-Christians, is primarily composed of individuals who are Christians and/or Westerners. Conversely, the other group consists of those who belong to other religions, such as Islam and Eastern religions, with its mode being the cross-category Asia-Pacific-Islam. Finally, the cluster tree is obtained (F).
\begin{table}[b]
\caption{\label{tab:illustrative} Cross frequencies of $462$ subjects with respect to their religious belief and geographic area of origin (fictitious data for illustrative purposes) are shown on the left. On the right, the table presents the estimated densities $\delta$ for each cross-category, as described in Equation \ref{eq:mutual_info}.}
\centering
\begin{tabular}{cc}
\begin{tabular}{lccc}
\toprule
& Christianity & Islam & Eastern Religions \\
\midrule
Europe & 85 & 5 & 1 \\
America & 110 & 2 & 3 \\
Africa & 20 & 30 & 5 \\
Asia-Pacific & 1 & 120 & 80 \\
\bottomrule
\end{tabular}
&
\begin{tabular}{ccc}
\toprule
\multicolumn{3}{c}{$\delta$} \\
\midrule
0.780 & 0.000 & 0.069 \\
1.000 & 0.024 & 0.037 \\
0.061 & 0.065 & 0.012 \\
0.007 & 0.866 & 0.766 \\
\bottomrule
\end{tabular}
\end{tabular}
\end{table}

%Consequently, the number of clusters is directly determined by the procedure, which does not require this information to be specified a priori, unlike other methods for clustering categorical data in the literature.
%Similar to \cite{stuetzle2010generalized}, we aim to find a single path for each node rather than having multiple paths connecting the same point. The MST is used to ensure that among all the possibile paths, the one with the highest weights is selected, so that the density is maximised.

\section{Discussion}
\label{modal_discussion}

The proposed method represents an ambitious attempt to extend modal clustering, jointly with its main strengths and flexibility, to the setting of categorical data. In fact, while the purpose has required to loosen some formal concepts, the proposed procedure claims the introduction of a circumscribed notion of clusters for categorical data, along with the operational steps to identify automatically the composition and number of clusters. These steps can be interpreted in the same guise as the steepest ascent and level set based methods usually employed in the continuous counterpart.  
In this section we discuss in more depth some aspects about the proposed procedure, to provide insights on its behaviour, the link with the inherent literature and some limitations. 

\begin{itemize}
%\item  \emph{mmm} The MC-C method offers several notable advantages. First and foremost, its definition of clusters for categorical data is well-defined and aligns closely with natural intuition. Operationally, this method extends modal clustering techniques, originally developed for continuous data, to categorical variables. To the best of our knowledge, this is the first density-based approach for clustering categorical data that does not require to specify in advance the number of clusters. Instead, the number of clusters is automatically determined based on the inherent association structure within the data.
\item \emph{Clusters and association 
between variables}. When clustering continuous data, targeting a formally defined concept of clusters aids the choice of the method and the subsequent evaluation of the resulting partition, according to some basic principle of inference. This is usually prevented in the categorical setting, where, to the best of our knowledge, the notion of clusters is usually imposed by the clustering method only implicitly.
Additionally, the idea to account for the association between variables %considering it as a crucial aspect for clustering categorical data. Although, a preliminary idea that leverages the discrepancy between observed and expected frequencies  was proposed by 
 makes the proposed notion close to some natural intuition. Considering the association among variables as an aspect to evaluate the clusterability is in fact somewhat shared by further approaches, yet never stated explicitly so far. Indeed, \citet*{ganti1999cactus} proposes an heuristic method which leverages on some measure of discrepancy between observed and expected frequencies under the hypothesis of independence.   
 Also Latent Class Analysis implicitly considers the association of variables when clustering data, by assuming independence among variables conditional to the cluster membership. However, this assumption is not always realistic, prompting \citet*{marbac2015model} to propose an extension that allows variables to be grouped into inter-dependent and intra-dependent categories to account for intra-class correlation, thereby partially considering this idea of association.
 Similar to LCA, MC-C defines clusters to capture the dependency structure conditionally on group membership. Specifically, we leverage the association between variables to obtain a clustering structure in which the dependencies between variables are minimised, as these dependencies are captured within the clusters themselves. Consequently, conditional to the clustering structure, the resulting contingency tables exhibit reduced or null levels of association.
\item \emph{Number of variables.} While the procedure has been illustrated for the easiest case of two categorical variables, the outlined concepts and measures  extend to an arbitrary number of variables. When computed across multiple variables, mutual information is referred to as multi-information. For a set of $p$ categorical variables, $X_1, \dots, X_p$, the multi-information is defined by the equation
\begin{equation}\label{eq:multi-info}
	I(X_1, \dots, X_p) = \left(\sum_{i = 1}^p H(X_i)\right) - H(X_1, \dots, X_p).
\end{equation}
For simplicity, consider the three-variable scenario where the categorical variables $A$, $B$, and $C$ have respectively $(A_1,\ldots, A_M)$, $(B_1, \ldots, B_L),$ and $(C_1, \ldots, C_S)$ categories. The individual contribution to the mutual information for a generic cross-category $(A_m, B_l, C_s)$ will then be
\begin{equation}
\label{eq:ind_contr_multi}
	\delta(A_m, B_l, C_s) = P(A_m, B_l, C_s)\log\left[\frac{P(A_m, B_l, C_s)}{P(A_m) P(B_l) P(C_s)}\right],
\end{equation}
which works as a density measure in the same guise as the two-variables case.
It is also worthwhile to note that, working on the contingency table instead of the observations, and depending on the number of observed variables and cross-categories, as done by MC-C, may results in a relevant computational savings.  

\item \emph{Density estimates.} In section \ref{sec:novelmethod} an extended notion of density for categorical data is proposed, based on the normalised individual contribution to mutual information. It stems to reason that the true involved probabilities are not known and must be estimated. Maximum likelihood estimation by means of log-linear models is a natural choice, leading to estimate both the joint probabilities and the expected ones under the hypothesis of independence via the corresponding empirical counterparts.  
However, possible limitations may arise when dealing with complex and high-dimensional datasets. Here, with the term \textit{high-dimensional} we refer to scenarios where the number of observed levels for each variable is large, resulting in a high number of observed cross-categories. In these scenarios, data sparsity, defined as the presence of very low observed frequencies across cross-categories, becomes a significant concern for two main reasons. On one hand, the estimates of the expected probabilities under the assumption of independence, may be neither robust nor reliable. On the other hand, data sparsity can lead to a disaggregation of information across the various levels of the variables, making the association structure more difficult to be unveiled. 
Consequently, this can diminish the algorithm’s ability to detect meaningful relationships, potentially resulting in suboptimal clustering outcomes or over granular clustering results.
A possible workaround to mitigate these issues, as suggested by \citet{simonoff1995smoothing} is to provide smoothed estimates. 
For instance, the kernel estimator proposed by \citet{li2003nonparametric}, based on \citet{aitchison1976multivariate}'s method, estimates joint probability distributions for categorical data without splitting the sample into cells in finite-sample applications, as it is typically done by the frequency estimator.  
%The estimator is defined as
%\begin{equation}
%\label{eq:li_racine}
%\widehat{p}(x) = \frac{1}{N} \sum_{j = 1}^N L(X_j, x; \lambda) 
%\end{equation}
%where $X$ is the vector of $\sum p\cdot M_i$ dummy variables with $p$ being the number of categorical variables and $M_i$ the number of categories of each categorical variable. For the $p$ categorical variables, each having $M_i$ categories, $p$ vectors $Z_i$ with $i = 1, \dots, p$ are obtained, with each vector comprising $M_i$ dummy variables. Therefore we have that
%\begin{equation}
%	L(Z, z; \lambda) = \prod_{i=1}^p \prod_{m=1}^{M_i} l(Z_{im}, z_m) = (1 - %\lambda)^{M - g_{mz}} \lambda^{g_{mz}}
%\end{equation}
%where $g_{mz}$ represents the number of disagreement components between $Z_{im}$ and $z_m$, and $\lambda \in \{0, 1/n\}$ is the smoothing parameter that is generally estimated from the data. However, in cases of sparse data, a more appropriate bandwidth can be specified to yield higher frequencies for cross-categories, addressing sparsity issues. As the bandwidth increases, the frequencies in the contingency table gradually become more evenly spread, leading to a shift toward a situation of independence and producing less granular clustering results. 

%Future research could focus on developing dimensionality reduction techniques or feature selection methods based on mutual information.
\item \emph{One cluster or many clusters?} As already mentioned, independence among variables shall be intended as lack of clusterability. Whereas in the continuous case the lack of clusters corresponds to an underlying unimodal density, in the categorical setting we may equivalently intend it as the aggregation of cross categories in a single ``unimodal'' group or the presence of as many groups as the number of cross-categories. For this reason, whenever the procedure should result in oversegmenting the data at hand, some caution should be used in interpreting results, since this situation might in fact be indicator of a general lack of clusterability. A good practice, for this reason, can be the one of evaluating the goodness of fit of a null log-linear model on the data, before proceeding with clustering. 
\item \emph{Further operational aspects.} Another operational aspect to consider is how to handle ties in the process of finding the maximal spanning tree. Specifically, it is necessary to determine a single path for each node when multiple paths have the same weight.
In the rare event that this occurs, we analyse the number of shared levels between the nodes. For example, if the node $(A_1, B_1, C_1)$ is connected to both $(A_1, B_1, C_2)$ and $(A_1, B_2, C_2)$ with the same weight, we retain only the link between $(A_1, B_1, C_1)$ and $(A_1, B_1, C_2)$ since they share a higher number of categories. Ties are broken at random in case the number of common categories is equal too.  
\end{itemize}

\section{Some numerical studies}
\label{numerical_studies}
\subsection{Simulated data}

The simulation study presented herein aims to evaluate the effectiveness and the performance of the proposed method, both in absolute terms and in comparison to some competitors. %With the main focus of investigating the proposed method ability to identify group structures when categorical are at hand, a wide range of simulations with different settings has been carried out. 
Of major concern is assessing the impact that different sample sizes and association structures of the data have on the results.

Generating categorical data for clustering purposes can be considered somehow more challenging than simulating such data for other purposes. The difficulty arises due to the lack of a clear and universally accepted definition of group. This complexity is further amplified when it is required to simulate categorical data that interact with each other, exhibit different levels of dependency, and include clusters with varying degrees of separation. For this reason, to avoid potential biases in the results that could arise from customizing a simulation study, we rely on one already designed to assess other methods, i.e. the work conducted by \citet{aschenbruck2020cluster}, as it presents some consistency with the notion of cluster considered here.

Originally designed for mixed variables, the study serves as a source to evaluate clustering also when dealing with categorical data only. In particular, this specification is designed to construct scenarios in which categorical data exhibit  some degree of association, as measured by a parameter $\theta \in [0.5, 1],$ with a value of $0.5$ indicating complete independence among variables, and a value of $1$ representing perfect dependency. 
The simulation outline is summarized in Table \ref{tab:as_simstudy} and further detailed in \citet{aschenbruck2020cluster}. 
We consider $5$ scenarios with progressively increasing $\theta$ values, each reflecting a different level of association among variables. Each scenario assumes the existence of $K = 2$ clusters, with observations evenly distributed among them. 
For each setting, $500$ samples of size $200$ and $1000$ are generated.  

In all settings MC-C is compared with other competing clustering approaches introduced in Section \ref{related_works}, including dissimilarity-based clustering methods such as PAM and K-modes, and model-based clustering methods like LCA and clustMD. All these approaches require the specification of the number of clusters in advance, which is fixed at the true number of clusters generated.

To measure the agreement between the actual labeling and the reconstructed results, we rely on the Fowlkes–Mallows index \citep{fowlkes1983method}. This index ranges from 0, indicating the method's ineffective ability to lead to the generated partition, to 1, reflecting a perfect classification, as it is directly proportional to the number of true positives. In fact, while the most widespread measure to compare partitions is the Adjusted Rand Index \citep{hubert1985comparing}, this provides misleading results when individuals are clustered into a single group, returning zero even when compared with a strongly imbalanced true labeling and the misclassification is in fact low. Unlike the ARI, the Fowlkes–Mallows index provides a more nuanced evaluation, avoiding potential biased and skewed results in certain scenarios.

\begin{table}[t]
\centering
\caption{Summary of the simulation study in \citet{aschenbruck2020cluster} for categorical data.}
\label{tab:as_simstudy}
\begin{tabular}{lcccc}
\toprule
Variable association & \#Clusters & \makecell{\#Categorical \\ variables} & \makecell{\#Categories for \\ each variable} & \makecell{Sample size \\ for each cluster} \\ 
\midrule
Very low ($\theta = 0.6$) & 2 & 2 & 2 & 100, 500\\ 
Low ($\theta = 0.7$)  & 2 & 2 & 2 & 100, 500\\ 
Medium ($\theta = 0.8$) & 2 & 2 & 2 & 100, 500\\ 
High ($\theta = 0.9$)  & 2 & 2 & 2 & 100, 500\\ 
Very high ($\theta = 0.95$) & 2 & 2 & 2 & 100, 500\\  
\bottomrule
\end{tabular}
\end{table}

\begin{figure}[tb]
    \centering
    \begin{subfigure}[c]{0.45\textwidth}
        \includegraphics[width=\textwidth]{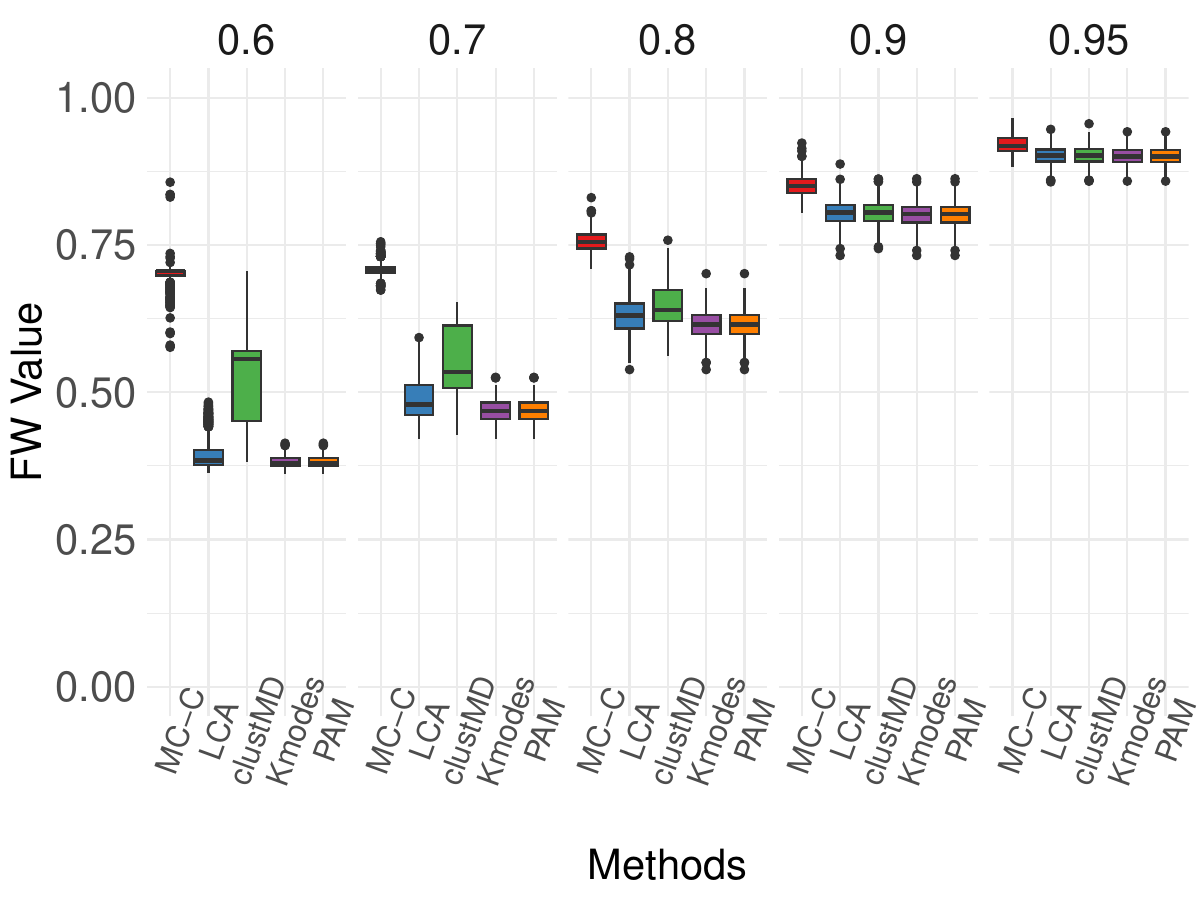}
    \end{subfigure}
    \begin{subfigure}[c]{0.45\textwidth}
        \includegraphics[width=\textwidth]{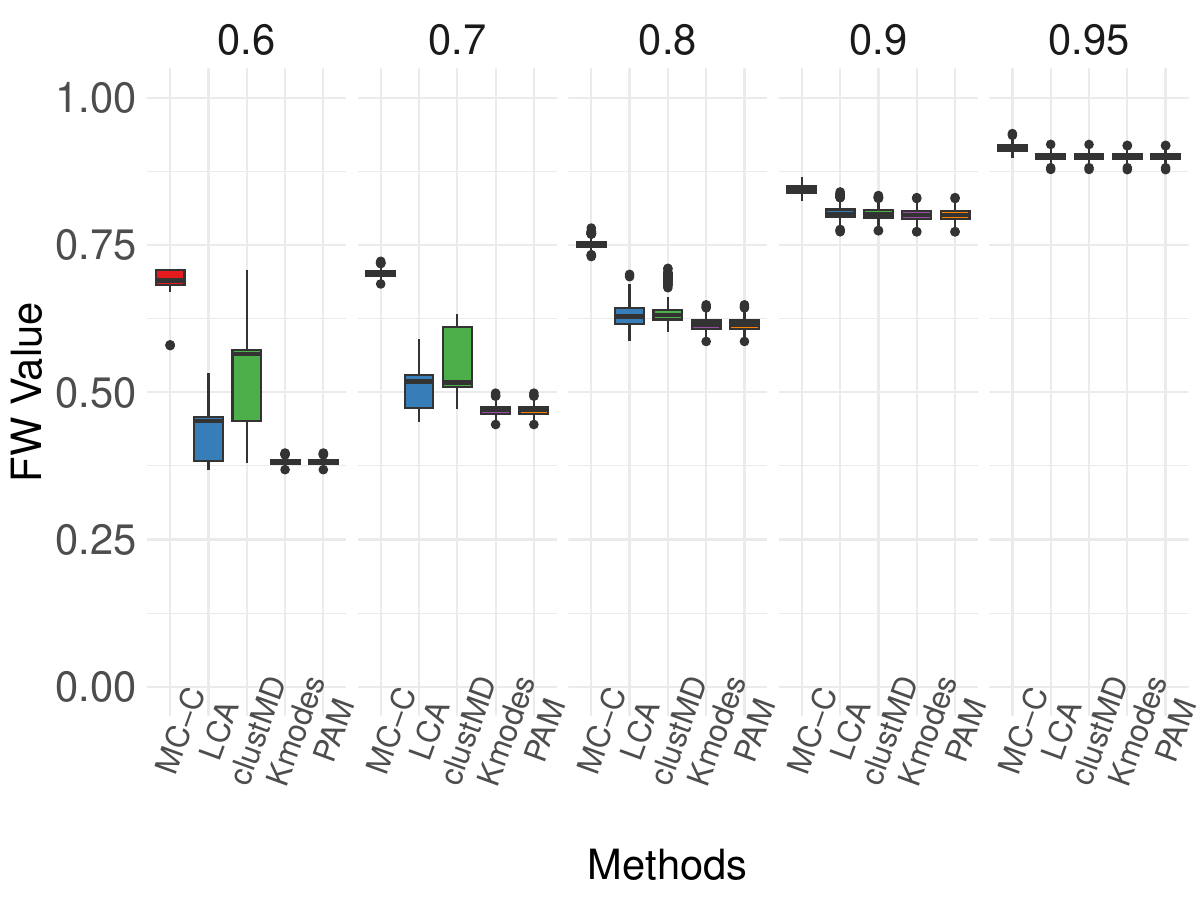}
    \end{subfigure}
    \caption{Boxplots of the Fowlkes-Mallows index between the true and the estimated clustering results for the MC-C method and competing methods. Results are shown for sample sizes of 100 (left) and 500 (right) per cluster, and across five levels of $\theta$.}
    \label{fig:catas20}
\end{figure}

The results of the simulation study, found in Figure~\ref{fig:catas20}, show that the MC-C method outperforms the competitors, with the FW Index that consistently favours it across all scenarios. As expected, clustering accuracy improves with larger sample sizes and increasing values of the parameter $\theta$. This result holds across all the considered methods, suggesting that the association between variables provides insights about data clusterability even to the competing methods. Notably, MC-C shows robustness even under challenging conditions, i.e. when variables are almost independent ($\theta = 0.6$). In contrast, other methods experience a more drastic decrease, with some reaching FW index below $0.5$ in worst-case scenarios. It is worth noting that in this simulation study, the competing methods have an advantage as they are set up to determine the correct number of groups; in contrast, MC-C identifies this information automatically, which makes the results even more meaningful.

\subsection{Real data}

In this section, we present two illustrative applications of our proposed methodology using publicly available real world data.  %Table~\ref{tab:deviance} that reports the deviance of the null log-linear model defined as
%\begin{equation*}
	%\log(N_{ml}) = \lambda + \lambda_m^A + \lambda_l^B
%\end{equation*}
%for the generic two variables case.
Here again we compare our approach with LCA, clustMD, K-modes, and PAM. Unlike the simulation study, here a ground truth labeling is not available, hence we estimate this number using automatic procedures. The Bayesian Information Criterion (BIC) serves as criterion for the LCA and clustMD methods. For the PAM and K-modes methods, we rely respectively on the Average Silhouette Width index (ASW), that measures the tradeoff between similarity of observations in the same cluster and dissimilarity of observations in different clusters, and the elbow method which considers the sum of the within-cluster simple-matching distance for each cluster. 

In the first example we analyse the renown Titanic dataset \citep{dawson1995unusual} reporting information on economic status (class), sex and survival on the $2201$ passengers involved in the infamous disaster of $1912$. 
The variables present a strong association (deviance of the null log-linear model equal to $1018.32$ with $10$ degrees of freedom). %Indeed, it is sadly known how being male and economically disadvantaged had a negative impact on the survival rate.

\begin{figure}[t!]
\centering
    \begin{subfigure}{.2\linewidth}
        \includegraphics[width=5cm, height=2.5cm, angle=-90]{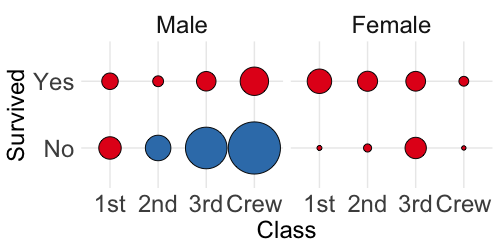}
        \caption{MC-C}
        \label{titanic_mcccm}
    \end{subfigure}
 %   \quad
    \begin{subfigure}{.19\linewidth}
        \includegraphics[width=5cm, height=2.5cm, angle=-90]{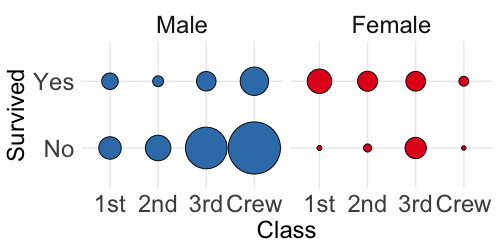}
        \caption{clustMD}
        \label{titanic_clustMD}
    \end{subfigure}
    \begin{subfigure}{.19\linewidth}
        \includegraphics[width=5cm, height=2.5cm, angle=-90]{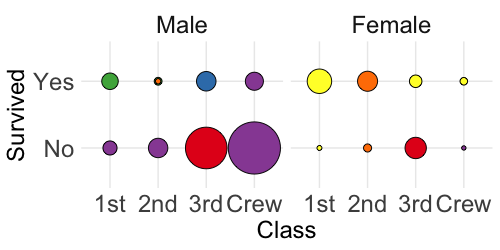}
        \caption{Kmode}
        \label{titanic_kmode}
    \end{subfigure}
%    \quad
    \begin{subfigure}{.18\linewidth}
        \includegraphics[width=5cm, height=2.5cm, angle=-90]{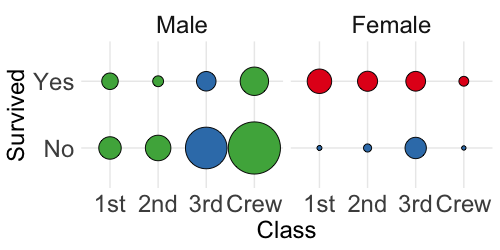}
        \caption{LCA}
        \label{titanic_lca}
    \end{subfigure}
    \begin{subfigure}{.18\linewidth}
    \includegraphics[width=5cm, height=2.5cm, angle=-90]{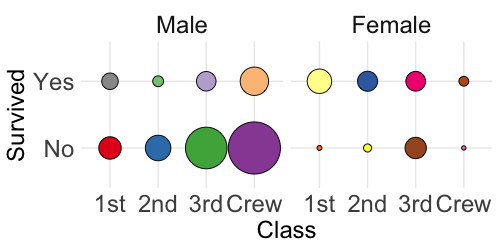}
        \caption{PAM}
        \label{titanic_pam}
    \end{subfigure}
    \caption{Graphical representation of the clustering results obtained from different methods for the Titanic dataset. Each plot visualises the contingency table, with the diameter of each circle proportional to the cell frequency and colours indicating the clustering assignments.}
    \label{fig:titanic_results}
\end{figure}

Results are reported in Figure \ref{fig:titanic_results}. Each sub-figure displays the clustering within a representation of the contingency tables used to visualise the data. 
Among the reference methods, PAM assigns a cluster to each cell in the contingency table, resulting in an obvious partition of the data. Similarly, LCA and K-modes cluster the data in three and six groups, respectively. ClustMD’s partition is driven by the \textit{gender} only while, in contrast, the MC-C method identifies two clusters that highlight the key information contained in the data. Specifically, one group consists solely of men from the second and third classes, as well as crew members, who are known to have been mostly penalised by the lack of lifeboats.

The second example revolves around $4526$ applicants to UC Berkeley graduate school across six large departments \citep*{bickel1975sex}. For each applicant, information on admission status and gender is recorded. The dataset is known as a notable example of the Simpson paradox, as while it seems that there is gender bias in the admission system that favours male students, a more detailed analysis at the department level shows a different perspective. Specifically, departments A and B have the highest admission rates, but only a few women applied to them. In contrast, more women applied to departments C, D, E, and F, which are the most competitive. Therefore, also in this example there is a strong association structure of the variables (deviance of the null log-linear model equal to $2097.67$ with $16$ degrees of freedom).

\begin{figure}[t!]
\centering
    \begin{subfigure}{0.19\textwidth}
       \includegraphics[width=2.5cm, height=4cm]{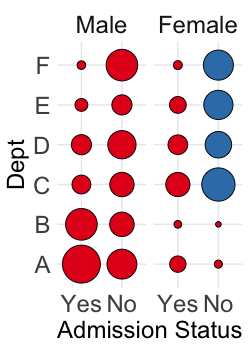}
        \caption{MC-C}
        \label{berkley_mcccm}
    \end{subfigure}
    \begin{subfigure}{0.19\textwidth}
       \includegraphics[width=2.5cm, height=4cm]{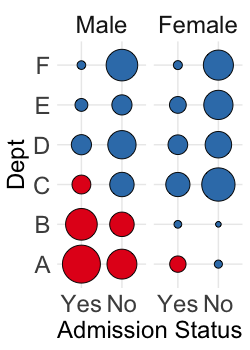}
        \caption{ClustMD}
        \label{berkley_clustMD}
    \end{subfigure}
    \begin{subfigure}{0.18\textwidth}
       \includegraphics[width=2.5cm, height=4cm]{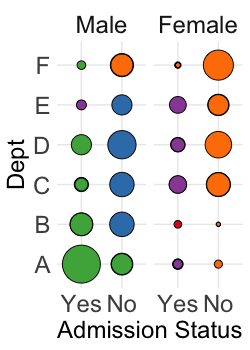}
        \caption{Kmode}
        \label{berkley_kmode}
    \end{subfigure}
    \begin{subfigure}{0.18\textwidth}
       \includegraphics[width=2.5cm, height=4cm]{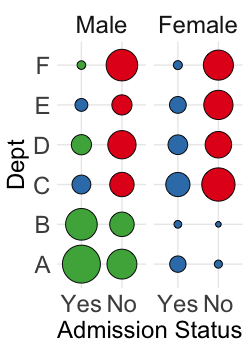}
        \caption{LCA}
        \label{berkley_lca}
    \end{subfigure}
    \begin{subfigure}{0.18\textwidth}
       \includegraphics[width=2.5cm, height=4cm]{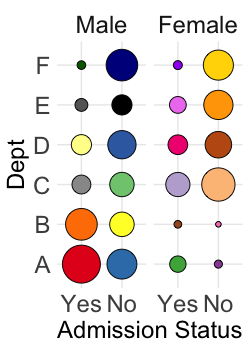}
        \caption{PAM}
        \label{berkley_pam}
    \end{subfigure}
    \caption{Graphical representation of the clustering results obtained from different methods for the Berkley dataset. Each plot visualises the contingency table, with the diameter of each circle proportional to the cell frequency and colours indicating the clustering assignments.}
    \label{fig:berkley_results}
\end{figure}

%------------------- Commento BERKLEY -------------------------%
Results are presented in Figure~\ref{fig:berkley_results}. 
PAM retrieves a number of clusters equal to the number of cells in the contingency table, yielding a non significant result. K-modes finds five clusters. LCA and clustMD successfully isolate the information related to the most selective departments but information related to the gender is missing. In contrast, the clusters identified by MC-C comprises all female applicants who were rejected in the more competitive departments (C, D, E, F) thus highlighting the source of potential bias in the admission process.

% With respect to the Berkley admission dataset, we present the clustering results in Figure~\ref{fig:berkley_results}. 
% The clusters identified by MC-C comprises all female applicants who were rejected in the more competitive departments (C, D, E, F) thus highlighting the source of potential bias in the admission process.
% When examining the results of the other clustering methods, we observe the that PAM retrieves a number of clusters equal to the number of cells in the contingency table, yielding a non significant result. LCA and clustMD successfully isolate the information related to the most selective departments but information related to the gender is missing. K-modes finds five clusters that lack of meaningful interpretation.

\section{Final remarks}
\label{conclusion}

When dealing with categorical data, our intuition for a new notion of clusters arises from observing the frequency patterns in contingency tables, which range from independent variables to variables that are strongly associated. On this ground, we introduced a novel notion of clusters that complies with a natural intuition and hinges on the twofold concept of high frequency and association between variables. Groups are defined as highly populated, with respect to the case of independence, aggregations of cross-categories of the observed variables. This concept aligns conceptually with the modal formulation of the clustering problem, which we have taken advantage of to borrow some operational
tools and build a new clustering method that jointly extends the ideas of connected sets,
gradient ascent, and density, typical of the non-parametric clustering setting.
Both simulations and real data applications yield promising results, suggesting that our method has strong potential for extending the non-parametric clustering approach to categorical data analysis. When comparing our method with existing approaches, the
results indicate that all methods perform satisfactorily and exhibit similar behaviour
for large sample sizes. However, in the most challenging clustering scenarios, with a low level of dependency between variables, the performance of the MC-C method is particularly noteworthy, as it outperforms its competitors. Moreover, contrary to existing approaches, this method does not require the specification of the number of clusters.\\

\noindent
\textbf{Funding} This research received no external funding.\\

\noindent
\textbf{Data Availability} The datasets analyzed herein are publicly available.\\

\section*{Declarations}

\textbf{Ethical Approval} This research contains no studies with human participation or animals performed by any
authors.\\

\noindent
\textbf{Conflict of Interest} The authors declare no competing interests.

\bibliography{sn-bibliography}% common bib file
\end{document}